\documentclass[3p]{elsarticle}
\usepackage{ecrc}

\volume{}

\firstpage{1}

\runauth{Varghese and Buyya, Next Generation Cloud Computing}

\jid{FGCS}

\usepackage{amssymb}

\usepackage[figuresright]{rotating}
\usepackage[hyphens]{url}
\usepackage{float}
\usepackage{color}
\usepackage{afterpage}
\usepackage{pdflscape}

\begin{document}

\begin{frontmatter}

\dochead{Accepted to Future Generation Computer Systems, 07 September 2017}

\title{Next Generation Cloud Computing:\\New Trends and Research Directions\footnote{This paper can be cited as follows: {\sffamily Blesson Varghese and Rajkumar Buyya, ``Next Generation Cloud Computing: New Trends and Research 	Directions,'' Future Generation Computer Systems, ISSN: 0167-739X, Elsevier Press, Amsterdam, The Netherlands, 2017 (in press).}}}

\author{Blesson Varghese\corref{mycorrespondingauthor}}
\address{School of Electronics, Electrical Engineering and Computer Science, Queen's University Belfast, UK}
\cortext[mycorrespondingauthor]{Corresponding author}
\ead[url]{www.blessonv.com}
\ead{varghese@qub.ac.uk}

\author{Rajkumar Buyya}
\address{Cloud Computing and Distributed Systems (CLOUDS) Laboratory\\School of Computing and Information Systems, The University of Melbourne, Australia}
\ead[url]{www.buyya.com}
\ead{rbuyya@unimelb.edu.au}

\begin{abstract}
The landscape of cloud computing has significantly changed over the last decade. 
Not only have more providers and service offerings crowded the space, but also cloud infrastructure that was traditionally limited to single provider data centers is now evolving. 
In this paper, we firstly discuss the changing cloud infrastructure and consider the use of infrastructure from multiple providers and the benefit of decentralising computing away from data centers. 
These trends have resulted in the need for a variety of new computing architectures that will be offered by future cloud infrastructure. 
These architectures are anticipated to impact areas, such as connecting people and devices, data-intensive computing, the service space and self-learning systems. 
Finally, we lay out a roadmap of challenges that will need to be addressed for realising the potential of next generation cloud systems. 
\end{abstract}


\end{frontmatter}

\section{Introduction}
\label{sec:introduction}
\afterpage{
  \begin{landscape}

\begin{figure}
		\includegraphics[scale=0.962]{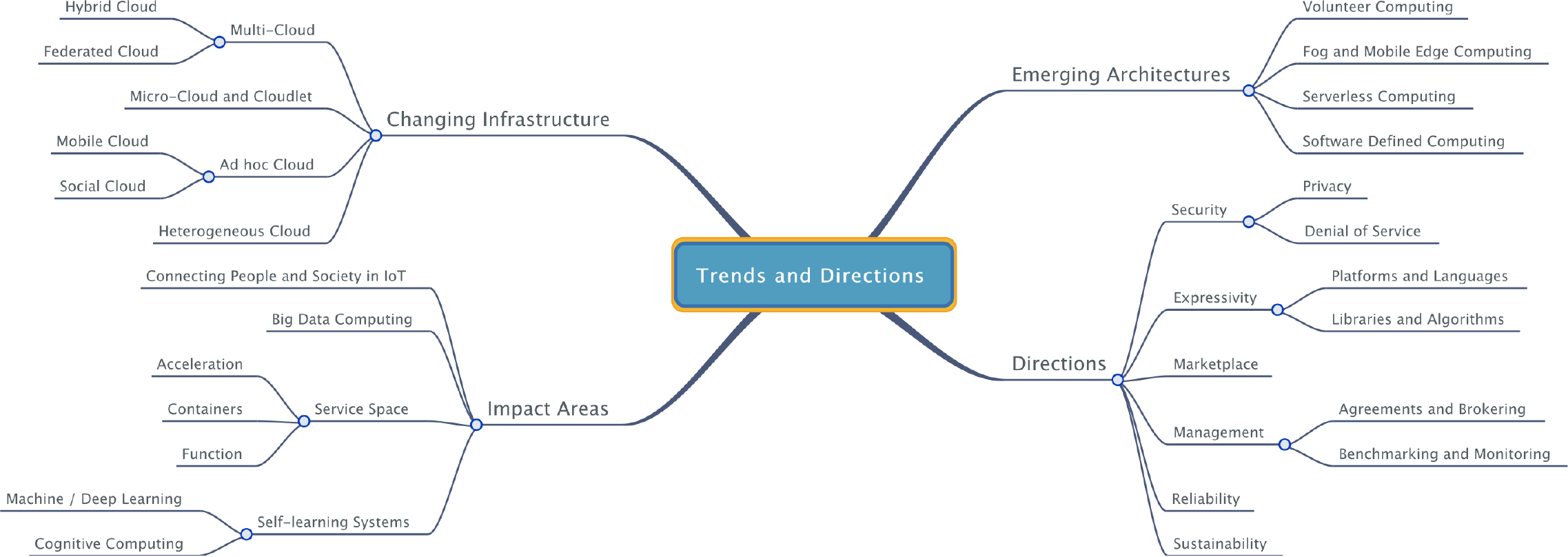}
	\centering
	\caption[Caption]{A snapshot of trends and directions in next generation cloud computing}
	\label{fig:figure1}
\end{figure}
\end{landscape}
}

Resources and services offered on the cloud have rapidly changed in the last decade. These changes were underpinned by industry and academia led efforts towards realising computing as a utility~\cite{vision-1}. This vision has been achieved, but there are continuing changes in the cloud computing landscape which this paper aims to present.

Applications now aim to leverage cloud infrastructure by making use of heterogeneous resources from multiple providers. This is in contrast to how resources from a single cloud provider or data center were used traditionally. Consequently, new computing architectures are emerging. This change is impacting a number of societal and scientific areas. In this discussion paper, we consider \textit{`what future cloud computing looks like'} by charting out trends and directions for pursuing meaningful research in developing next generation computing systems as shown in Figure~\ref{fig:figure1}. 

The remainder of this paper is organised as follows. 
Section~\ref{sec:infrastructure} presents a discussion of the evolving infrastructure on the cloud. 
Section~\ref{sec:computing} highlights the emerging computing architectures and their advantages.
Section~\ref{sec:impact} considers a number of areas that future clouds will impact.
Section~\ref{sec:direction} sets out a number of challenges that will need to be addressed for developing next generation cloud systems. 
Section~\ref{sec:conclusions} concludes this paper.

\section{Changing Infrastructure}
\label{sec:infrastructure}
The majority of existing infrastructure hosting cloud services comprises dedicated compute and storage resources located in data centers. 
Hosting cloud applications on data centers of a single provider is easy and provides obvious advantages. However, using a single provider and a data center model poses a number of challenges. A lot of energy is consumed by a large data center to keep it operational. Moreover, centralised cloud data centers like any other centralised computing model is susceptible to single point failures. Additionally, data centers may be geographically distant from its users, thereby requiring data to be transferred from its source to resources that can process it in the data center. This would mean that applications using or generating sensitive or personal data may have to be stored in a different country than where it originated. 

\begin{figure}
	\centering
	\includegraphics[width=0.95\textwidth]{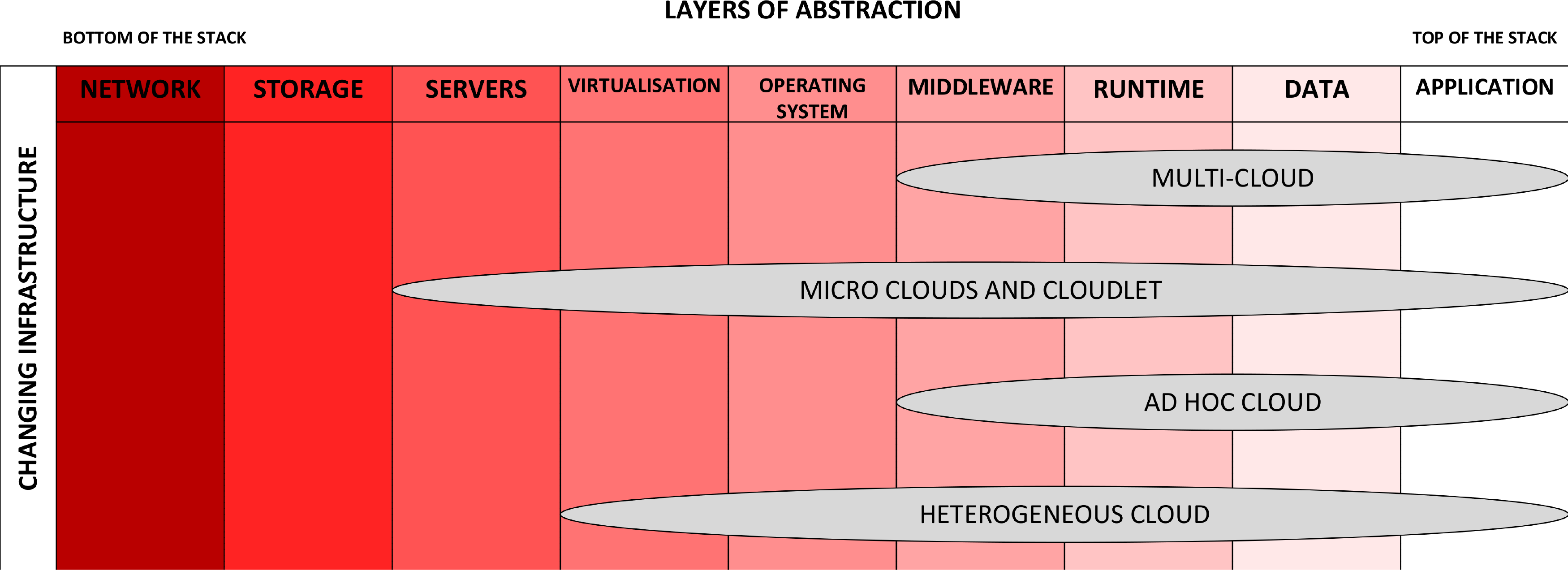}
	\caption{Layers of abstraction in the cloud stack that need to accommodate changes brought about by the evolving infrastructures} 
	\label{fig:figure2}
\end{figure}

{\color{black}
Strategies implemented to mitigate failures on the cloud include using redundant compute systems in a data center, multiple zones and back up data centers in individual zones. 
}
However, alternate models of using cloud infrastructure instead of using data centers from a single provider have been proposed in recent years~\cite{infrastructure-1}. In this paper, we consider the multi-cloud, micro cloud and cloudlet, ad hoc cloud and heterogeneous cloud to demonstrate the trends in changing infrastructure of the cloud. The feasibility of these have been reported in literature and will find real deployment of workloads in next generation cloud computing. {\color{black}Figure~\ref{fig:figure2} shows the different layers of the cloud stack where changes need to be accommodated due to the evolving infrastructure. We consider nine layers of abstraction that contribute to the cloud stack, namely network (bottom of the stack), storage, servers, virtualisation, operating system, middleware, runtime, data and application (top of the stack). For facilitating multi-cloud environments and ad hoc clouds, changes will be required from the middleware layer and upwards in the stack. Heterogeneous clouds can be achieved with changes two further layers down the stack from the virtualisation layer. Micro clouds and cloudlet infrastructure may require re-design of the servers that are employed and therefore changes are anticipated from the server layer.} 

{\color{black}
We note that there has also been significant changes in the area of data storage on the cloud over the last decade. There are at least three levels of abstraction provided with respect to data storage~\cite{datastorage-1}. At the block level, direct attached storage, such as Amazon EC2, App Engine and Azure VM, and block storage, such as EBS, Blob Storage and Azure Drive, are available for providing quickest access to data for VMs. At the file level, object storage, such as Amazon S3, Google Storage, and Azure Blob, and online drive storage, such as Google and Sky drives, are commonly available services. At the database level, relational data-stores, such as Google Cloud SQL and SQL Azure Blob, and semi-structured data storage, such as Simple DB, Big Table and Azure Table, are available. However, in this paper the focus is on how computing on the cloud has changed over the last decade.
}

\subsection{Multi-cloud}
{\color{black}The traditional notion of multi-cloud was leveraging resources from multiple data centers of a provider. Then applications were hosted to utilise resources from multiple providers~\cite{multicloud-1,multicloud-4}.}
Rightscale estimates that current businesses use an average of six separate clouds\footnote{\url{http://www.forbes.com/sites/joemckendrick/2016/02/09/typical-enterprise-uses-six-cloud-computing-services-survey-shows/#e2207a47be31}}. 

The use of multi-clouds are increasing, but there are hurdles that will need to be overcome. For example, common APIs to facilitate multi-cloud need to account for different types of resources offered by multiple providers. {\color{black}This is not an easy given that more resources are rapidly added to the cloud marketplace and there are no unified catalogues that report a complete set of resources available on the cloud.} Further, the abstractions, including network and storage architectures differ across providers, {\color{black}which makes the adoption of multi-cloud bespoke to each application rather than using a generic platform or service. Along with the different resources, hypervisors, and software suites employed, the pricing and billing models are significantly different across providers, all of which results in significant programming effort required for developing a multi-cloud application. All management tasks, such as fault tolerance, load balancing, resource management and accounting need to be programmed manually since there are no unifying environments that make these possible. Examples of APIs that alleviate some of these challenges include Libcloud\footnote{\url{https://libcloud.apache.org/}} and jClouds\footnote{\url{https://jclouds.apache.org/}}. However, further research is required for enabling adoption of clouds across multiple providers.
}

\textit{Hybrid Cloud:} 
A multi-cloud can take the form of a hybrid cloud - a combination of public and private clouds or a combination of public and private IT infrastructure~\cite{hybridcloud-1,hybridcloud-2}. These clouds cater for bursty demands or resource demands known beforehand.  
The benefit of using hybrid clouds for handling sensitive data is known~\cite{hybridcloud-5}. 
It is estimated that 63\% of organisations using the cloud have adopted a hybrid cloud approach\footnote{\url{http://www.cloudpro.co.uk/cloud-essentials/hybrid-cloud/6445/63-of-organisations-embracing-hybrid-environments}} with use-cases reported in healthcare\footnote{\url{https://usa.healthcare.siemens.com/medical-imaging-it/image-sharing/image-sharing-archiving-isa}} and energy sectors\footnote{\url{http://w3.siemens.com/smartgrid/global/en/products-systems-solutions/smart-metering/emeter/pages/emeter-cloud.aspx}}. 
The key challenge in setting up a hybrid cloud is network related. Bandwidth, latency and network topologies {\color{black} will need to be considered for accessing a public cloud from a private cloud~\cite{hybridcloud-10}. Network limitations can result in an ineffective hybrid cloud. Dedicated networking between clouds may enable more effective infrastructure, but requires additional management of private resources, which can be a cumbersome task.}

\textit{Federated Cloud:}
There are a number of benefits in bringing together different cloud providers under a single umbrella resulting in a federated cloud~\cite{federatedcloud-1,federatedcloud-2}. This can provide a catalogue of services and resources available as well as makes applications interoperable and portable. The EU based EGI Federated Cloud is an example of this and brings together over 20 cloud providers and 300 data centers\footnote{\url{https://www.egi.eu/}}. Federated clouds can address the vendor lock-in problem in that applications and data can be migrated from one cloud to another. This is not easy given different abstractions, resource types, networks and images as well as variable vendor specific costs for migrating large volumes of data. {\color{black}In addition, from a business and organisational perspective it may be easier for smaller providers to federate to expand their reach and extend their services. However, larger providers may not be willing to federate with other providers given that they have geographically spread resources. }
{\color{black}More recently, there are ongoing efforts to federate resources that are located outside cloud data centers, which is considered next.} 

\subsection{Micro cloud and Cloudlet}
{\color{black}Data centers occupy large amounts of land and consume lots of electricity to provide a centralised computing infrastructure. This is a less sustainable trend, and alternate low power and low cost solutions are proposed. There are recent efforts to decentralise computing towards the edge of the network for making computing possible closer to where user data is generated~\cite{microcloud-0}.} Small sized, low cost and low power computing processors co-located with routers and switches or located in dedicated spaces closer to user devices, referred to as micro clouds are now developed for this purpose~\cite{microcloud-2,microcloud-3,microcloud-4}. {\color{black} However, there are no public deployments given the challenges in networking micro cloud installations over multiple sites. In the UK, there are efforts to connect micro clouds for general purpose computing\footnote{\url{http://gow.epsrc.ac.uk/NGBOViewGrant.aspx?GrantRef=EP/P004024/1}}.}

Micro clouds lend themselves in reducing latency of applications and minimising the frequency of communication between user devices and data centers. Odroid boards\footnote{\url{http://www.hardkernel.com}} and Raspberry Pis\footnote{\url{https://www.raspberrypi.org/}} for example are used to develop micro clouds. However, integration of micro clouds to the existing computing ecosystem is challenging and efforts are being made in this direction~\cite{osmotic-1}. {\color{black}One of the key challenges is scheduling applications during run time to make use of micro clouds along with a data center. This includes partitioning an application and its data across both high end and low power processors to improve the overall performance measured by user-defined objectives. In a decentralised cloud computing approach, application tasks will need to be offloaded both from data centers and user devices on to micro clouds. The challenge here is in using micro clouds (that may or may not be always available) with network management abstraction between the cloud and the edge without depending on the underlying hardware.}

The aim of a cloudlet is similar to a micro cloud in extending cloud infrastructure towards the edge of the network~\cite{cloudlet-1,cloudlet-2}, but is used in literature in the context of mobile computing. It is used for improving the latency and overall Quality of Service (QoS) of mobile applications. Next generation computing systems will integrate computing on the cloudlet to service local traffic and reduce network traffic towards cloud data centers beyond the first hop in the network. The Elijah\footnote{\url{http://elijah.cs.cmu.edu/}} project is an example of advances in the cloudlet arena. 

\subsection{Ad hoc cloud}
{\color{black} The use of micro clouds and cloudlets will need to leverage on the concept of ad hoc computing that has existed from the grid era. For example, SETI@home\footnote{\url{http://setiathome.berkeley.edu/}} was a popular project that aimed to create a computing environment by harnessing spare resources from desktops using BOINC\footnote{\url{http://boinc.berkeley.edu/}}. 
The concept of ad hoc clouds is based on the premise of ad hoc computing in that underutilised resources, such as servers owned by organisations can be harnessed to create an elastic infrastructure~\cite{adhoccomputing-1,adhoccomputing-2}. This is in contrast to existing cloud infrastructure which is largely data center based and in which the resources available are known beforehand. 

However, the context of an ad hoc cloud is changing with increasing connectivity of a large variety of resources to the cloud~\cite{adhocnew-3}. This is becoming popular for smaller mobile devices, such as smartphones~\cite{mobileadhoccomputing-1,mobileadhoccomputing-2}, which on an average have less than a 25\% per hour of usage~\cite{adhocnew-1,adhocnew-2}. The spare resources of smartphones can contribute to creating an ad hoc infrastructure (such as cloudlets) that supports low latency computing for non-critical applications in public spaces and transportation systems. The assumption here is that one device is surrounded by a large number of devices that will complement computing for the former device. Although such an infrastructure is not reliable, it may be used in conjunction with existing data centers to enhance connectivity. {\color{black}Reliability of ad hoc clouds can be interpreted in two ways. Firstly, ad hoc clouds will need robust job and resource management mechanisms to surmount malicious activity (this problem was noted on desktop Grids)~\cite{adhoc-robust-1}. Secondly, the battery of mobile devices that contribute to an ad hoc cloud will be drained~\cite{adhoc-robust-2}. Hence, mobile devices that complement computing will need to be volunteer devices or devices that are commandeered in hostile environments (for example, a terrorist attack in a city) or for critical applications (for example, during a natural or man-made disaster).}
However, such ad hoc clouds may be an enabler for deployments of cloudlets that improve the QoS of applications. 
}

\subsection{Heterogeneous Cloud}
{\color{black}Heterogeneity in cloud computing can be considered in at least two ways. The first is in the context of multi-clouds, in which platforms that offer and manage infrastructure and services of multiple cloud providers are considered to be a heterogeneous cloud. Heterogeneity arises from using hypervisors and software suites from multiple vendors.

The second is related to low-level heterogeneity at the infrastructure level, in which different types of processors are combined to offer VMs with heterogeneous compute resources. In this paper, the latter is referred to as heterogeneous clouds. While supercomputers have become more heterogeneous by employing accelerators, such as NVIDIA GPUs or Intel Xeon Phis, cloud data centers mostly employ homogeneous architectures~\cite{heterogeneouscloud-1}. More recently heterogeneous cloud data center architectures have been proposed\footnote{\url{http://www.harness-project.eu/}}.} In the vendor arena, Amazon along with other providers offer GPU-based VMs, but accelerators are not yet fully integrated into the computing ecosystem. {\color{black}This is because it is not yet possible for a programmer to fully develop and execute code oblivious to the underlying hardware. There are a number of efforts in this direction, but the key challenge is achieving a high-level abstraction that can be employed across multiple architectures, such as GPUs, FPGAs and Phis~\cite{heterogeneouscloud-2,heterogeneouscloud-3,heterogeneouscloud-4,heterogeneouscloud-5}}.
Further applications that already execute on the cloud cannot be scheduled onto heterogeneous resources. Efforts in this direction are made by the CloudLightning\footnote{\url{http://cloudlightning.eu/}} project. The concept of a heterogeneous cloud may extend beyond the data center. For example, ad hoc clouds or microclouds could be heterogeneous cloud platforms.

\section{Emerging Computing Architectures}
\label{sec:computing}
Conventional cloud computing requires applications to simply follow a two tier architecture\footnote{\url{https://cloudacademy.com/blog/architecting-on-aws-the-best-services-to-build-a-two-tier-application/}}
{\color{black}In a two tier architecture front end nodes, such as user devices, make use of a service offered by the cloud. The entire business logic and database logic are located in the cloud.} With the increasing number of sensor rich devices, such as smartphones, tables and wearables, large volumes of data is generated. Gartner forecasts that by 2020 over 20 billion devices will be connected to the internet consequently generating 43 trillion gigabytes of data\footnote{\url{http://www.gartner.com/newsroom/id/3165317}}. 
This poses significant networking and computing challenges that will degrade the Quality-of-Service (QoS) and Experience (QoE) that cannot be met by existing infrastructure. Adding more centralised cloud data centers or eliminating them from the computing system will not address the problem. Instead a fundamentally different approach extending the computing ecosystem beyond cloud data centers towards the user will pave the way forward. This will include resources at the edge of the network or resources voluntarily contributed by owners, which is typically not considered in conventional cloud computing. 

\begin{figure}
	\centering
	\includegraphics[width=0.95\textwidth]{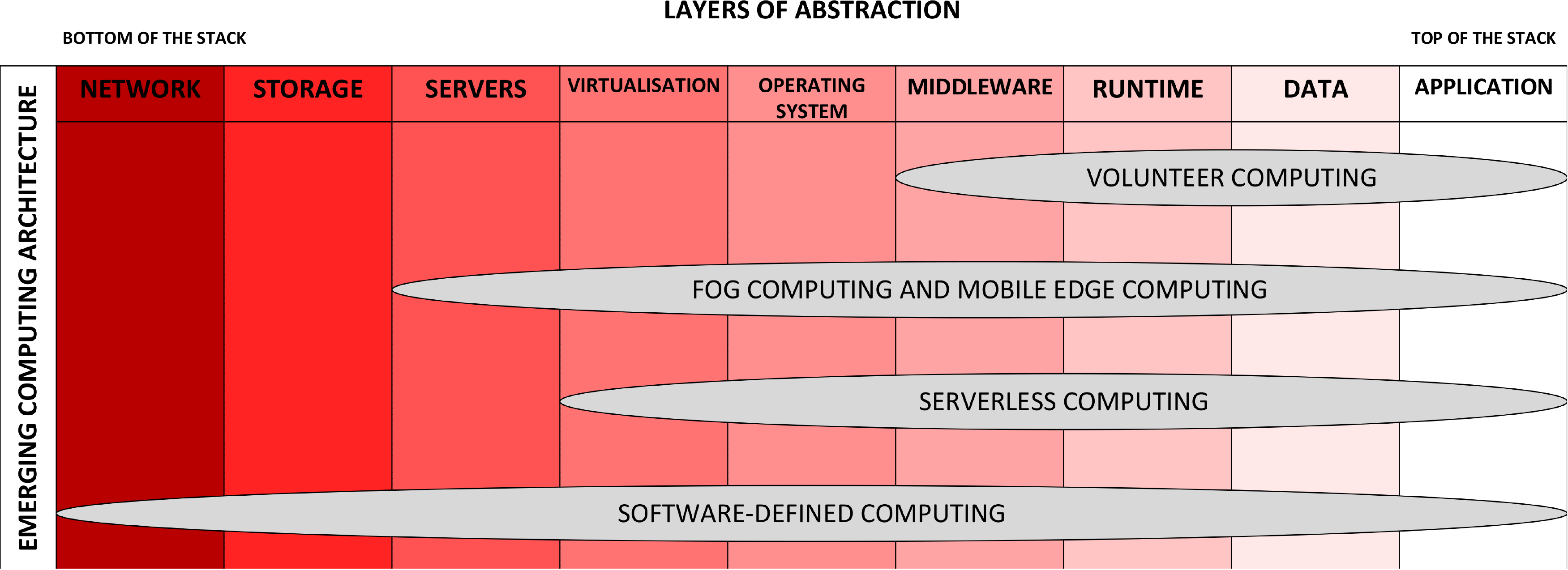}
	\caption{Layers of abstraction in the cloud stack that need to accommodate changes brought about due to emerging computing architectures} 
	\label{fig:figure3}
\end{figure}

The cloud computing infrastructure is evolving and requires new computing models to satisfy large-scale applications. In this paper, we consider four computing models, namely volunteer computing, fog and mobile edge computing, serverless computing and software-defined computing that will set trends in future clouds. Figure~\ref{fig:figure3} shows the different layers of the cloud stack where changes need to be accommodated for the emerging computing architectures.

\subsection{Volunteer Computing}
{\color{black}Ad hoc clouds and cloudlets are emerging to accommodate more innovative user-driven and mobile applications that can benefit from computing closer to user devices. The availability of compute resources is not guaranteed in an ad hoc cloud or cloudlet as in a conventional data center and therefore a pay-as-you-go or an upfront payment for reserving compute, storage or network resources will not be suitable. Instead, a crowd funded approach in which spare resources from user computers or devices are volunteered for creating an ad hoc cloud. Such a computing model may be used to support applications that have a societal or scientific focus. }


Volunteer cloud computing~\cite{volunteercloud-1,volunteercloud-2,volunteercloud-3} can take different forms. For example, users of a social network may share their heterogeneous computing resources in the form of an ad hoc cloud. This is referred to as `social cloud computing'~\cite{socialcloud-1,socialcloud-2}. More reliable owners are rewarded through a reputation marker within the social network. The Cloud@Home project rewards volunteers by payment for their resource donations~\cite{cloudhome-1}. Gamification is also reported as an incentive~\cite{cloudgamification-1}. Similar research is also reported in the name of `peer-to-peer cloud computing'~\cite{peertopeercloud-1}.

The challenges that need to be overcome to fully benefit from volunteer cloud computing will be firstly in minimising the overheads for setting up a highly virtualised environment given that the underlying hardware will be heterogeneous and ad hoc. Moreover, there are security and privacy concerns that will need to be addressed to boost confidence in the public to more readily become volunteers for setting up ad hoc clouds. Furthermore, a consistent platform that can integrate social networks with cloud management will need to emerge.

\subsection{Fog and Mobile Edge Computing}
The premise of fog computing is to leverage the existing compute resources on edge nodes, such as mobile base stations, routers and switches, {\color{black}or integrate additional computing capability to such (network) nodes along the entire data path between user devices and a cloud data center. The general characteristic of these nodes are that they are resource constrained.} This will become possible if general purpose computing can be facilitated on existing edge nodes or additional infrastructure, such as micro clouds or cloudlets are deployed. Preliminary research reports the applicability of fog computing for use-cases, such as in online games and face recognition~\cite{facerecognition-1}. {\color{black}The obvious benefits of using fog computing includes minimising application latency and improving the Quality-of-Service (QoS) and Experience (QoE) for users while leveraging hierarchical networking and tapping into resources that are traditionally not employed for general purpose computing. Therefore, it is anticipated that fog computing may enable realising the Internet-of-Things (IoT) vision. We note that fog computing will not make centralised clouds obsolete, but will work in conjunction with them to facilitate more distributed computing.}

One characteristic of using fog computing is that applications can be vertically scaled across different computing tiers. This will enable only essential data traffic beyond the data source. {\color{black}Workloads can be offloaded from cloud data centers on to edge nodes~\cite{offload-1,offload-2} or from user devices on to edge nodes~\cite{offload-3,offload-4} to process data near its source rather than in geographically distant locations. Additionally, an aggregating model~\cite{aggregation-1,aggregation-2}, in which data is aggregated from multiple devices or sensors will be possible.} Application and operating system containers may substitute the more heavyweight virtual machines for deploying workloads. 

The term `Mobile Edge Computing (MEC)' is used in literature~\cite{mec-1,mec-2}, which is similar to fog computing in that the edge of the network is employed. However, it is limited to the mobile cellular network and does not harness computing along the entire path taken by data in the network. In this computing model the radio access network may be shared with the aim to reduce network congestion. Application areas that benefit include low latency content delivery, data analytics and computational offloading~\cite{mecoffload-1,mecoffload-2} for improving response time.
Intel has reported the real life use of MEC\footnote{\url{https://builders.intel.com/docs/networkbuilders/Real-world-impact-of-mobile-edge-computing-MEC.pdf}} and industry led proof-of-concept models that support MEC have been developed\footnote{\url{http://mecwiki.etsi.org/index.php?title=Main_Page}}. It is anticipated that MEC will be adopted in 4G/5G networks\footnote{\url{http://www.etsi.org/images/files/ETSIWhitePapers/etsi_wp11_mec_a_key_technology_towards_5g.pdf}}.

To realise fog computing and MEC at least two challenges will need to be addressed. Firstly, complex management issues related to multi-party service level agreements~\cite{sla-1,sla-2}, articulation of responsibilities and obtaining a unified platform for management given that different parties may own edge nodes. Secondly, enhancing security and addressing privacy issues when multiple nodes interact between a user device and a cloud data center~\cite{fogsecurity-1,fogsecurity-2}. The Open Fog consortium\footnote{\url{https://www.openfogconsortium.org/}} is making a first step in this direction.

\subsection{Serverless Computing}
Conventional computing on the cloud requires an application to be hosted on a Virtual Machine (VM) that in turn offers a service to the user. If a web server is hosted on a cloud VM, for example then the service owner pays for the entire time the server application is hosted (regardless of whether the service was used).
{\color{black}The metrics against which the performance of an application is generally benchmarked include latency, scalability, and elasticity. Therefore, development efforts on the cloud focus on these metrics. The cost model followed is `per VM per hour' and does not take idle time into account (the VM was provisioned, but the server was idle since there were no requests or the application was not running). This is because the VM on which the server is running requires to be provisioned. However, with decentralised data center infrastructure that may have relatively less processing power, it will not be ideal to continually host servers that will remain idle for a prolonged period of time. Instead an application in a fog or MEC environment may be modularised with respect to the time taken to execute a module or the memory used by the application. This will require a different cost model that accounts for the memory consumed by the application code for the period it was executed and the number of requests processed
}

As the name implies `serverless' does not mean that computing will be facilitated without servers~\cite{newserverless-3,newserverless-4}. In this context, it simply means that a server is not rented as a conventional cloud server and developers do not think of the server and the residency of applications on a cloud VM\footnote{\url{https://d0.awsstatic.com/whitepapers/AWS_Serverless_Multi-Tier_Architectures.pdf}}. From a developers perspective challenges such as the deployment of an application on a VM, over/under provisioning of resources for the application, scalability and fault tolerance do not need to be dealt with. The infrastructure, including the server is abstracted away from the user {\color{black} and instead properties, such as control, cost and flexibility are considered.} 

In this novel approach, functions (modules) of the application will be executed when necessary without requiring the application to be running all the time~\cite{newserverless-1,newserverless-5}. Sometimes this is also referred to as Function-as-a-Service or event-based programming~\cite{newserverless-7}. An event may trigger the execution of a function or a number of functions in parallel. Examples of platforms that currently support this architecture includes AWS Lambda\footnote{\url{https://aws.amazon.com/lambda/}}, IBM OpenWhisk\footnote{\url{https://developer.ibm.com/openwhisk/}} and Google Cloud Functions\footnote{\url{https://cloud.google.com/functions/}}. {\color{black}The open source project OpenLambda aims to achieve serverless computing~\cite{newserverless-2}. There are implementations of using serverless on the edge, referred to as Lambda@Edge\footnote{\url{http://docs.aws.amazon.com/lambda/latest/dg/lambda-edge.html}}. The benefit of using serverless computing for scientific workflows has been demonstrated~\cite{newserverless-6}.}

Forbes predicts that the use of serverless computing will increase given that billions of devices will need to be connected to the edge of the network and data centers\footnote{\url{http://www.forbes.com/sites/ibm/2016/11/17/three-ways-that-serverless-computing-will-transform-app-development-in-2017/?cm_mc_uid=24538571706014848428726&cm_mc_sid_50200000=1484842872\#4f04f02565a3}}. {\color{black} It will not be feasible to have idle servers in resource constrained environments.
The challenges that will hinder the widespread adoption of serverless computing will be the radical shift in the properties of an application that a programmer will need to focus on; not latency, scalability and elasticity, but those that relate to the modularity of an application, such as control and flexibility. Another challenge is developing programming models that will allow for high-level abstractions to facilitate serverless computing. The effect and trade-offs of using traditional external services along with serverless computing services will need to be investigated in orchestrating future cloud-based systems.}

\subsection{Software-Defined Computing}
There is a large amount of traffic that traditionally did not exist in a two-tier cloud architecture. This is due to the ever increasing number of devices that are catered for by the Internet. Consequently, there is an increasing volume of data that needs to be transferred from one location to another to support applications that rely on multiple cloud services. To efficiently manage this, the networking technology needs to support a dynamic architecture. \textit{Software Defined Networking (SDN)} is an approach of isolating the underlying hardware in the network from the components that control data traffic~\cite{sdn-1,sdn-2}. This abstraction allows for programming the control components of the network to obtain a dynamic network architecture. 

{\color{black}In the context of future clouds, there are a number of challenges and opportunities relevant to developing SDN. Firstly, there are challenges in developing hybrid SDNs in lieu of centralised or distributed SDNs~\cite{sdn-3}. Research is required to facilitate physically distributed protocols while logically centralised control tasks can be supported. The second challenge is in developing techniques to capture Quality-of-Service by taking both the network and cloud infrastructure into account~\cite{sdn-3}. This is required for capturing end-to-end QoS and improving user experience in both virtualised network and hardware environments. Thirdly, the interoperability of Information-Centric Networking (ICN) will need to be facilitated as cloud networks adopt ICN over SDN~\cite{sdn-3}. Fourthly, developing mechanisms for facilitating network virtualisation for different granularities, for example per-job or per-task granularity~\cite{fgcs-sdn-1}.
}

With emerging distributed cloud computing architectures, `software defined' could be applied not only to networking, but also to storage\footnote{\url{https://www.opensds.io/}} and compute as well as resources beyond data centers for delivering effective cloud environments~\cite{sdcc-1,sdcc-2}. This concept when applied to compute, storage and networks of a data center and resources beyond is referred to as Software Define Computing (SDC). This will allow for easily reconfiguring and adapting physical resources to deliver the agreed QoS metrics. The complexity in configuring and operating the infrastructure is alleviated in this case.

\section{Avenues of Impact}
\label{sec:impact}
\begin{figure}
	\centering
	\includegraphics[width=0.95\textwidth]{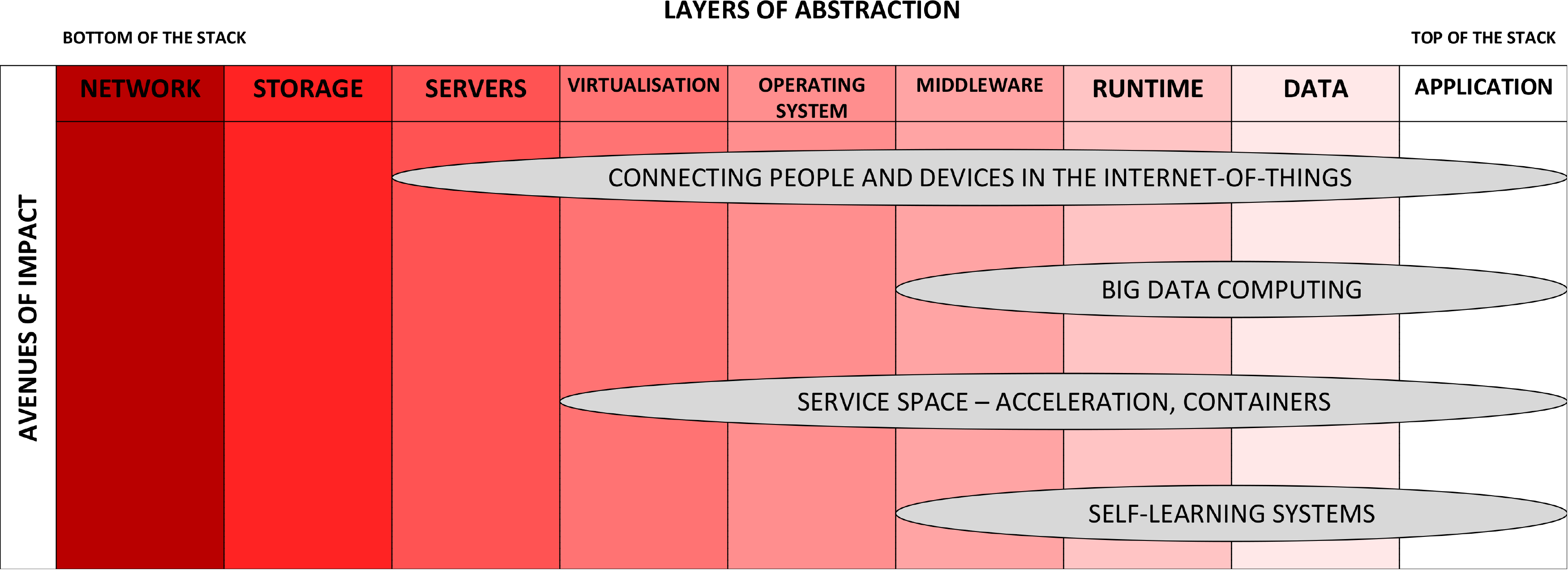}
	\caption{Layers of abstraction in the cloud stack that need to accommodate changes brought about due to the avenues of impact} 
	\label{fig:figure4}
\end{figure}

Next generation cloud computing systems are aimed at becoming more ambient, pervasive and ubiquitous given the emerging trends of distributed, heterogeneous and ad hoc cloud infrastructure and associated computing architectures. This will impact at least the following four areas considered in this paper. Figure~\ref{fig:figure4} shows the different layers of the cloud stack where changes need to be accommodated for the avenues of impact.

\subsection{Connecting People and Devices in the Internet-of-Things}

Innovation in the cloud arena along with prolific growth of the sensors and gadgets is bringing people, devices and the associated computing closer. 
The concept of combining multiple sensor environments, including sensors embedded into infrastructure (transportation, communication, buildings, healthcare and utilities) and sensors on user devices, wearables and appliances has resulted in the upcoming \textit{Internet of Things (IoT)}~\cite{iot-1,iot-2}. The aim is to improve the accuracy and efficiency of an actuation process and reduces human intervention. The `things' in the IoT context may vary from microchips, biometric sensors, sensors on mobile phones and electrical gadgets at home to sensors embedded on infrastructure monitoring pollution, temperature, light etc. 
Conventional cloud computing architectures will be limiting in connecting billions of things, but a combination of computing architectures presented in Section~\ref{sec:computing} will facilitate the IoT vision.

{\color{black}Among many, a key challenge will be related to end-to-end security in networks given that sensor networks, wireless networks, RFID devices, cloud data centers, edge nodes, public and private clouds will need to be integrated for achieving IoT systems. Current mechanisms in securing networks involves encryption and authentication, which will prevent outsider attacks. However, additional techniques are required against insider malicious attacks. This will require the development of secure reprogrammable protocols that allow the authentication of events that triggers a function in the network (as in serverless computing), thereby preventing malicious installations.}

{\color{black}Additionally, innovation in  sensing and decision-making is required. Traditionally, sensing involved physical sensors that are integrated in the environment, but future IoT systems will involve the integration of physical sensors, people-centric sensors (low cost sensing of the environment localised to the user) and human sensors (people provide data about their environment)~\cite{iot-3,iot-4,fgcs-smartcity-1}. The key challenge here is that data is unstructured and platforms that account for this are required and currently not available.}

\subsection{Big Data Computing}
{\color{black}
The consequence of emerging computing models is that they generate large volumes of data, referred to as `Big Data'. 
Data generated by organisations or users are transferred to a data store on the cloud (this is a result of employing a centralised `application to data center' architecture). The data that is stored may never be used again and is often referred to as dark data. It is usually expensive to move data out of the store and perform any analytics. The opportunity to process data is before it is stored in the cloud. 

As the cloud infrastructure becomes decentralised, more opportunities unveil to facilitate processing closer to where it is generated before storing it. For example, edge nodes may be used for processing image or video data before it is stored. However, existing research in big data usually considers centralised cloud architectures or multiple data centers. To leverage distributed cloud architectures there are a number of challenges that will need to be addressed.

Firstly, data processing and resource management on distributed cloud nodes\footnote{\url{http://www.cloud-council.org/deliverables/CSCC-Deploying-Big-Data-Analytics-Applications-to-the-Cloud-Roadmap-for-Success.pdf}}~\cite{bdc-2}. Whether they be ad hoc clouds, heterogeneous clouds or distributed clouds, there needs to be platforms that can take into account the ad hoc nature of nodes that may process data in a distributed cloud setting, the heterogeneity of processors and platforms that scale from low power processors to high-end processors without significant programming efforts. 

Secondly, building models for analytics that scale both horizontally and vertically. Current models typically scale horizontally across multiple nodes in a data center or across nodes in multiple data centers. In the future, models that scale vertically from low end processors to data center nodes will need to be developed. 

Thirdly, software stacks for end-to-end processing~\cite{bdc-1}. This relates to both the first and second challenges. Currently, most big data solutions assume a centralised cloud as the compute resource, but integrating micro clouds, cloudlets or traffic routing edge nodes in the software stack will need to be addressed. 
}

{\color{black}In the real-world, the volume of unstructured data as opposed to structured data is increasing. Often unstructured data is interconnected (for example, generated from social networks) and takes the form of natural language. One key challenge is achieving accurate and actionable knowledge from unstructured data. To address this challenge, one approach will be to transform unstructured data to structured networks and then to knowledge, referred to as data-to-network-to-knowledge (D2N2K)\footnote{\url{https://www.nsf.gov/awardsearch/showAward?AWD_ID=1705169\&HistoricalAwards=false}}. However, this is challenging since automated and distant supervision methods will need to be firstly developed~\cite{bigdatanew-1,bigdatanew-2}. Then methods will be required to derive knowledge from structured networks represented as graphs. Similarly, there are a number of challenges when performing analytics on large graphs. They include the need for designing novel mechanisms for fast searching and querying~\cite{bigdatanew-3,bigdatanew-5} and secure searching and indexing underpinned by homomorphic algorithms~\cite{bigdatanew-4}.
These remain open areas of research~\cite{bigdatanew-6,bigdatanew-7}.}

{\color{black}There is another dimension to the big data computing, which is the legal issues surrounding sensitive data generated by cloud applications. There are numerous legislative and regulatory constraints that aim to account for data access control, protection, privacy and security\footnote{\url{https://d0.awsstatic.com/whitepapers/compliance/Cloud_Computing_and_Data_Protection.pdf}}. For example in the European Union (EU) the transfer of personal data to non-member countries is forbidden, unless an adequate level of protection is ensured\footnote{\url{http://eur-lex.europa.eu/LexUriServ/LexUriServ.do?uri=CELEX:31995L0046:en:HTML}}. Providers such as Microsoft are ensuring Healthcare Insurance Portability and Accountability (HIPAA) compliance for its Azure cloud\footnote{\url{https://www.microsoft.com/en-us/TrustCenter/Compliance/HIPAA}} and Amazon are acknowledging EU data protection legislation in structuring its Amazon Web Services\footnote{\url{https://d0.awsstatic.com/whitepapers/compliance/Using_AWS_in_the_context_of_Common_Privacy_and_Data_Protection_Considerations.pdf}}.
}

\subsection{Service Space}
The abstraction of infrastructure, platforms and software were initially offered as services (IaaS, PaaS and SaaS) on the cloud. However, the service space is becoming richer with a wide variety of services. For example, to offer acceleration provided by GPUs to applications \textit{Acceleration-as-a-Service (AaaS)} has been proposed~\cite{aaas-1}. In the future, as more applications make use of hardware accelerators the AaaS space is expected to become more mature. Currently, GPU virtualisation technologies, such as rCUDA facilitate the use of GPU services~\cite{rcuda-1,rcuda-2}. {\color{black} However, most AaaS services still require applications to be specifically written for an accelerator. Further, a wider variety of accelerators, such as coprocessors, FPGAs and ASICs (such as Tensor Processing Units (TPUs) need to be integrated in future clouds to enable computing in device rich environments, such as fog computing and IoT.} 
There is ongoing research to migitate these challenges, for example the Anyscale Apps\footnote{\url{http://anyscale.org/}} project and the OpenACC initiative\footnote{\url{http://www.openacc.org/}}. 

Another area in the service space that is gaining significant traction is \textit{Container-as-a-Service (CaaS)}~\cite{caas-1,caas-2}. The benefits of deploying containers have been investigated for a variety of applications (although they are not applicable for all workloads). Consequently, containers are starting to be adopted as an alternate virtualisation technology. {\color{black} CaaS offers the deployment and management of containers, which will be required for workload execution in ad hoc clouds and micro clouds for enabling volunteer computing and fog computing, respectively}. Examples include Google Kubernetes\footnote{\url{https://kubernetes.io/}}, Docker Swarm\footnote{\url{https://www.docker.com/products/docker-swarm}} and Rackspace Carina\footnote{\url{https://getcarina.com/}}. However, avenues such as container monitoring and live migration will need to be developed~\cite{fgcs-migration-1}. Dealing with dependencies and the portability of containers remains an open issue. The security aspects of containers due to weak isolation relative to cloud VMs needs to be further understood.

With the adoption of event-based platforms for enabling serverless computing, more applications will make use of \textit{Function-as-a-Service (FaaS)}. The aim will be to execute functions on the cloud platform that are initiated by events. This is in contrast to current execution models in which an application is constantly running on the server to furnish a client request and is billed even when the server application remains idle when it is not servicing requests\footnote{\url{https://serverless.com/}}.   

\subsection{Self-learning systems}
Currently, a large volume of user generated data in the form of photo, audio and video and metadata, such as network and user activity information, are moved to the cloud. This is due to the availability of relatively cheaper data storage and back up on the cloud. There is ongoing research in applying machine learning to speech/audio recognition, text, image and video analysis, and language translation applications~\cite{cloudml-1}. This research is branded under the general umbrella of \textit{`Deep Learning'}~\cite{cloudml-2}. Traditionally machine learning algorithms were restricted to execution on large clusters given the large computational requirements. However, APIs and software libraries are now available to perform complex learning tasks without incurring significant monetary costs. Examples include the Google TensorFlow\footnote{\url{https://www.tensorflow.org/}} and Nervana Cloud\footnote{\url{https://www.nervanasys.com/cloud/}}. The availability of hardware accelerators, such as GPUs, in cloud environments has reduced the computing time for machine learning algorithms on large volumes of data~\cite{gpuml-1,gpuml-2,gpuml-3}. Interest from the industry in this area is due to the potential of deep learning in predictive analytics. 

A closely related avenue in the context of future clouds is \textit{Cognitive Computing}. In this visionary model, cognitive systems will rely on machine learning algorithms and the data that is generated to continually acquire knowledge, model problems and determine solutions. Examples include the use of IBM Watson for speech and facial recognition and sentiment analysis\footnote{\url{https://www.ibm.com/watson/}}. APIs and SaaS supporting Watson are currently available. The hardware employed in cognitive systems may rely on functions of the human brain and are inherently massively parallel (examples include the SyNAPSE~\cite{cc-1} and the SpiNNaker~\cite{cc-2} architectures). It is anticipated that these architectures will be integrated in next generation clouds.

\section{Research Directions}
\label{sec:direction}
\begin{figure}
	\centering
	\includegraphics[width=0.85\textwidth]{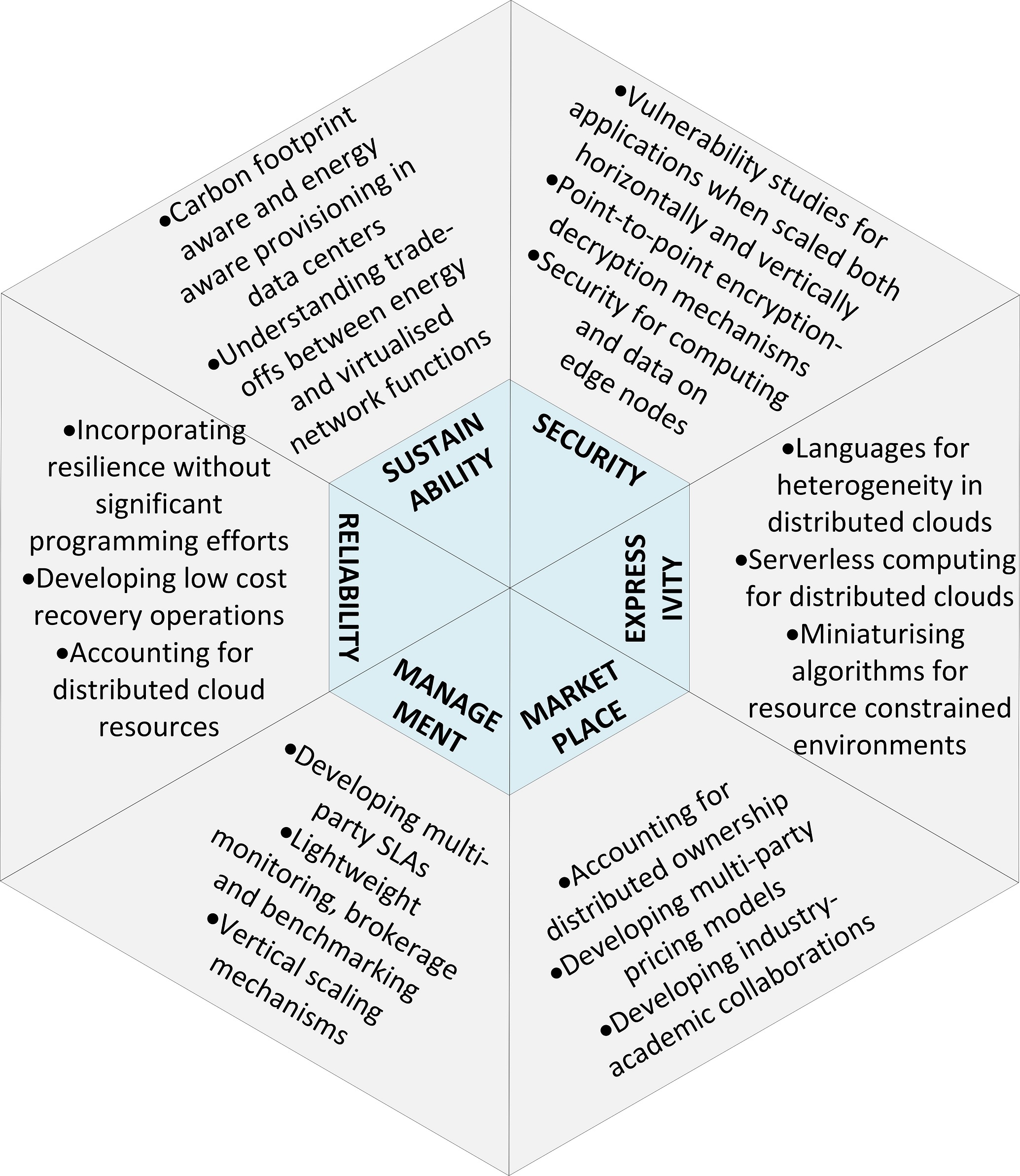}
	\caption{Research directions we present in this paper for next generation cloud computing} 
	\label{fig:figure5}
\end{figure}

In this section, we present a few directions (refer Figure~\ref{fig:figure5}) that academic cloud research can contribute to in light of the new trends considered in the previous sections.  

\subsection{Guaranteeing Enhanced Security}
The key to widespread adoption of computing remotely is security that needs to be guaranteed by a provider~\cite{cloudsecurity-1,cloudsecurity-2,fgcs-security-1}. In the traditional cloud, there are significant security risks related to data storage and hosting multiple users, which are mitigated by robust mechanisms to guarantee user and user data isolation. However, this becomes more complex, for example in the fog computing ecosystem, the above risks are of greater concern, since a wide range of nodes are accessible to users (for example, the security of traffic routed through nodes, such as routers~\cite{edgesecurity-1,edgesecurity-2}). For example, a hacker could deploy malicious applications on an edge node, which in turn may exploit a vulnerability that may degrade the QoS of the router. Such threats may have a significant negative impact. Moreover, if user specific data needs to be temporarily stored on multiple edge locations to facilitate computing on the edge, then privacy issues along with security challenges will need to be addressed. {\color{black}We recommend the design and development of methods to characterise and detect malwares at large scale~\cite{fgcs-mal-1}.}

{\color{black} Vulnerability studies that can affect security and privacy of a user when an application is scaled across both vertical (data centers, edge nodes, user devices) and horizontal (across multiple edge nodes and user devices) hierarchy will need to be considered. These studies will need to consider privacy concerns that are both inherited from traditional cloud systems and emerge from integrating sensors in the internet~\cite{fgcs-privacy-1}. One open area that will need to be considered for distributed clouds is the authentication of distributed (edge, peer, ad hoc, micro cloud) nodes. Another area is developing suitable encryption-decryption mechanisms that are less resource hungry and energy consuming that scale on resource deprived nodes of distributed clouds. Also, methods to detect intrusion, such as anomaly detection, will need to be designed for real-time resource and bandwidth limited environments for upcoming IoT workloads.
}  

Distributed Denial of Service (DDoS) has become a well known security threat in the cloud arena and is a potential threat with greater negative impact on distributed clouds~\cite{ddos-1,ddos-2}. Malicious users and hackers tend to exploit vulnerabilities of cloud services that are underpinned by virtualisation, auto-scaling mechanisms and multi-tenancy. Typically, the attack deprives other users on the cloud of resources and bandwidth, thereby making them incur more monetary cost for less optimised performance. This in turn negatively affects customer trust of the cloud provider and impacts cloud adoption. Often large-scale attacks that are reported in popular media have political and business motives. Three broad mechanisms, namely prevention, detection and mitigation are generally proposed to address such attacks. However, the development and adoption of concrete methods on the cloud are still in their infancy. The points of vulnerabilities increase as distributed cloud architectures are adopted and as more users and devices are connected to the cloud. Attack prevention, detection and mitigation methods will need to be further developed in conjunction with foundation technologies that enable the cloud.

\subsection{Achieving Expressivity of Applications for Future Clouds}
Expressing distributed applications using emerging computing models on changing infrastructure is an important research direction. 

Platforms and Languages: in addition to popular programming languages, there is a wide variety of services to deploy applications on the cloud. However, with the increasing emphasis on distributed cloud architectures there will be the need for developing platforms and toolkits that account for the integration and management of edge nodes. Given that distributed cloud applications will find its use-cases in user-driven applications, existing platforms cannot be used to easily program an application, such as a distributed workflow. The programming model that aims to exploit edge nodes will need to execute workloads on multiple hierarchical levels. Languages that support the programming model will need to take the heterogeneity of hardware and the capacity of resources in the workflow into account. If edge nodes available are more vendor specific, then the platforms supporting the workflow will need to account for it. This is more complex than existing models that make the cloud accessible.

{\color{black}
Platforms that facilitate serverless computing will need to be responsive in executing functions or microservices with limited start up latencies. Current delays in invoking functions are due to creating containers for each execution of a function. Although containers are faster than VMs existing container technology cannot be the unit of deployment. Alternate lean environments will be required to be integrated with platforms that facilitate Function-as-a-Service. 
}

Libraries and Algorithms: unlike large servers distributed cloud architectures will not support heavyweight software due to hardware constraints. For example, if a small cell base station with a 4-core ARM-based CPU
and limited memory is employed as an edge node in a distributed cloud model, then there is limited resources for executing complex data processing tools such as Apache Spark\footnote{\url{http://spark.apache.org}} that requires at least 8 cores CPU and 8 gigabyte memory for good performance. Here lightweight algorithms that can do reasonable machine learning or data processing tasks are required~\cite{kartakis2014real, santos2013dial}. Apache Quarks\footnote{\url{http://quarks.incubator.apache.org}}, for example, is a lightweight library that can be employed on small devices such as smart phones to enable real-time data analytics. However, Quarks supports basic data processing, such as filtering and windowed aggregates, which are not sufficient for advanced analytical tasks (e.g. context-aware recommendations). Machine learning libraries that consume less memory would benefit data analytics for edge nodes. 

{\color{black}
Current research is mostly targeted at developing platforms, libraries and languages for individual requirements of the emerging computing architectures. For example, individual software platforms are available for serverless computing or IoT. However, there are a number of common requirements for these emerging architectures and are not a design factor when developing platforms. Efforts towards developing a unified environment that can address the common requirements of emerging architectures to achieve interoperable and application independent environments will need to be a direction for research. Such unified environments can then be extended to suit individual requirements. We recommend the design and development of self-managing applications as a way forward for realising this~\cite{fgcs-selfmanaging-1}.
}
 
 \subsection{Developing a Marketplace for Emerging Distributed Architectures}
The public cloud marketplace is competitive and taking a variety of CPU, storage and communication metrics into account for billing~\cite{cloudpricing-1,cloudpricing-2}. For example, Amazon's pricing of a VM is based on the number of virtual CPUs and memory allocated to the VM. Distributed cloud architectures will require the development of a similar yet a more complex marketplace and remains an open issue. This will need to be developed with industry-academic collaborations (for example, the Open Fog consortium is set up with industry and academic partners to achieve open standards in the fog computing architecture). The marketplace will need to take ownership, pricing models and customers into account.

Typically, public cloud data centers are owned by large businesses. If traffic routing nodes were to be used as edge nodes in distributed cloud architectures, then their ownership is likely to be telecommunication companies or governmental organisations that may have a global reach or are regional players (specific to the geographic location. For example, a local telecom operator). Distributed ownership will make it more challenging to obtain a unified marketplace operating on the same standards. 

When distribution using the edge is considered, three possible levels of communication, which are between the user devices and the edge node, one edge node and another edge node, and an edge node and a cloud server, will need to be accounted in a pricing model. In addition, `who pays what' towards the bill has to be articulated and a sustainable and transparent economic model will need to be derived. The priority of applications executing on these nodes will have to be considered. If a serverless computing model is developed, then monitoring tools at the fine-grain level of functions will need to be designed. These are open research areas. 

Given that there are multiple levels of communication in emerging cloud architectures, there are potentially two customers. The first is an application owner running the service on the cloud who wants to improve the QoS for the application user. The second is the application user who could make use of a distributed architecture to improve the QoE when using a cloud service. For both the above, in addition to existing service agreements, there will be requirements to create agreements between the application owner, the nodes on to which an application is distributed and the user, which can be transparently monitored within the marketplace.

\subsection{Offering Efficient Management Strategies in the Computing Ecosystem}

On the cloud two key management tasks include (i) setting up agreements between parties involved and brokering services to optimise application performance, and (ii) benchmarking resources and monitoring services to ensure that high-level objectives are achieved. Typically, Service Level Agreements (SLAs) are used to fulfil agreements between the provider and the user in the form of Service Level Agreements (SLAs)~\cite{sla-1,sla-2}. This becomes complex in a multi-cloud environment~\cite{broker-4,broker-3} and in distributed cloud environments (given that nodes closer to the user could also be made accessible through a marketplace). If a task were to be offloaded from a cloud server onto an edge node, for example, a mobile base station owned by a telecommunications company, then the cloud SLAs will need to take into account agreements with a third-party. Moreover, the implications to the user will need to be articulated. The legalities of SLAs binding both the provider and the user in cloud computing are continuing to be articulated. Nevertheless, the inclusion of a third party offering services and the risk of computing on a third party node will need to be articulated. Moreover, if computations span across multiple edge nodes, then keeping track of resources becomes a more challenging task. 

Performance is measured on the cloud using a variety of techniques, such as benchmarking to facilitate the selection of resources that maximise performance of an application and periodic monitoring of the resources to ensure whether user-defined service level objectives are achieved~\cite{benchmarking-2,benchmarking-6,benchmarking-4,benchmarking-5}. Existing techniques are suitable in the cloud context since they monitor nodes that are solely used for executing the workloads~\cite{monitoring-0,monitoring-2}. On edge nodes however, monitoring will be more challenging, given the limited hardware availability. Benchmarking and monitoring will need to take into account the primary service, such as routing traffic, that cannot be compromised. Communication between the edge node and user devices and the edge node and the cloud and potential communication between different edge nodes will need to be considered. Vertical scaling along multiple hierarchical levels and heterogeneous devices will need to be considered. These may not be important considerations on the cloud, but becomes significantly important in the context of fog computing. {\color{black}The SLAs that are defined in future distributed clouds will need to implicitly account for security~\cite{fgcs-sla-1}.}

\subsection{Ensuring Reliability of Cloud Systems}
{\color{black}
Reliability of the cloud continues to remain a concern while adopting the cloud for remote computing and storage. Cloud failures have been reported affecting a number of popular services, such as DropBox and Netflix\footnote{\url{https://www.techflier.com/2016/01/25/top-20-high-profile-cloud-failures-all-time/}}$^{,}$\footnote{\url{http://www.nytimes.com/2012/12/27/technology/latest-netflix-disruption-highlights-challenges-of-cloud-computing.html?_r=1}}$^{,}$\footnote{\url{http://spectrum.ieee.org/computing/networks/understanding-cloud-failures}}. It is also reported that a 49-minute outage suffered by Amazon.com in 2013 cost the company more than \$4 million in lost sales\footnote{\url{http://tinyurl.com/jjkn235}}. As unplanned outages are inevitable, losses from outages will continue to escalate with the rapid growth of e-commerce businesses. 
Reliability becomes more challenging as the infrastructure becomes distributed. Recently, efforts are being made to design more reliable cloud data centers and services. 

On the infrastructure level, to deal with hardware failures due to targeted attacks and natural disasters, VMs and data are rigorously replicated in multiple geographic locations~\cite{resilientcomputing-1,resilientcomputing-2}. Proactive and reactive strategies so as to back up VMs taking network bandwidth and associated metrics are now inherent to designing cloud data centers. FailSafe is a Microsoft initiative for delivering disaster resilient cloud architectures which can be made use by a cloud application. However, incorporating resilient computing into distributed cloud applications remains challenging, still requires significant programming efforts and is an open area of research\footnote{\url{https://docs.microsoft.com/en-us/azure/guidance/guidance-resiliency-overview}}~\cite{fgcs-disaster-1}. Notwithstanding, disaster recovery is an expensive operation, and is required as a service to minimise recovery time and costs after a failure has occurred~\cite{resilientcomputing-3}.
Multi-cloud and multi-region architectures that scale both horizontally (geographically distributed) and vertically (not only in cloud data centers, but throughout the network) are recommended to avoid single points of failure~\cite{fgcs-geodistributed-1}. 
}

\subsection{Building Sustainable Infrastructure for the Future}
{\color{black}In 2014, it was reported that the US data centers consumed about 70 billion kilowatt-hours of electricity. This is approximately 2\% of the total energy consumed in the US\footnote{\url{http://www.datacenterknowledge.com/archives/2016/06/27/heres-how-much-energy-all-us-data-centers-consume/}}.
}
Data centers are huge investments which have adverse environmental impact due to large carbon footprints. While it may not be possible to eliminate data centers from the computing ecosystem, innovative and novel system architectures that can geographically distribute data center computing are required for sustainability. 

Useful contributions in this space can be achieved by developing algorithms that rely on geographically distributed data coordination, resource provisioning and carbon footprint-aware and energy-aware provisioning in data centers~\cite{sustainable-1,sustainable-5,sustainable-3,sustainable-4}. These will in turn minimise energy consumption of the data center and maximise the use of green energy while meeting an application's QoS expectations. {\color{black}Incorporating energy efficiency as a QoS metric has been recently suggested~\cite{sustainable-6}. This risks the violation of SLAs since VM management policies will become more rigorous aiming to optimise energy efficiency. However, there is a trade off between performance of the cloud resource and energy efficiency. This clearly is an open avenue for research. Intra and inter networking plays a key role in setting up efficient data centers. Virtualising network functions through software defined networking is an upcoming area to manage key services offered by the network. However, energy consumption is not a key metric that is considered in current implementations. An open area is the understanding of the trade off between energy consumption and network functions. Addressing this will provide insights into developing cloud infrastructure that are becoming more distributed.
}
Algorithms for application-aware management of power states of computing servers can be incorporated towards achieving more sustainable solutions in the long run. Moreover, methods that incorporate resilience in the event of outages and failures will be required. 

{\color{black}
Current Cloud systems primarily focus on consolidation of VMs to minimise energy consumption of servers. However, cooling systems and networks consume a significant proportion of the total energy consumed. Emerging techniques will need to be developed that manage energy efficiency of servers, networks and cooling systems. These techniques can leverage the interplay between IoT-enabled cooling systems 
and data center managers that dynamically make decisions on which resources to switch on/off in both time and space dimensions based on workload forecasts. 
}

\section{Summary}
\label{sec:conclusions}
\textit{So what does cloud computing in the next decade look like?} The general trend seems to be towards making use of infrastructure from multiple providers and decentralising computing away from resources currently concentrated in data centers. This is in contrast to traditional cloud offerings from single providers. Consequently, new computing models to suit the demands of the market are emerging. 

In this paper, we considered computing models that are based on voluntarily providing resources to create ad hoc clouds and harnessing computing at the edge of the network both for mobile and online applications. A computing model which will replace the traditional notion of paying for a cloud VM even when a server executing on the VM is idle was presented. The concept of integrating resilience and software-defined into distributed cloud computing is another emerging computing model that was highlighted in this paper.

Both the changing cloud infrastructure and emerging computing architecture will impact a number of areas. They will play a vital role in improving connectivity between people and devices to facilitate the Internet-of-Things paradigm. The area of data intensive computing will find novel techniques to address challenges related to dealing with volume of data. New services, such as containers, acceleration and function, is anticipated become popular. A number of research areas will find convergence with next generation cloud systems to deliver self-learning systems. 

These changes are being led both by the industry and academia, but there are a number of challenges that will need to be addressed in the future. In this paper we considered directions in enhancing security, expressing applications, managing efficiently and developing sustainable systems for next generation cloud computing. 

\section*{Acknowledgements}
We thank Professor Peter Sloot (Editor-in-Chief of FGCS) and all five anonymous reviewers for their suggestions and constructive comments that helped us to substantially improve the paper.

\bibliographystyle{elsarticle-num}
\bibliography{references}

\end{document}